# Effective Mass of a Migrating Interface

Xinyuan Song[1, 2], Chuang Deng[1*]

*1 Department of Mechanical Engineering, University of Manitoba, Winnipeg, MB R3T 2N2, Canada*

*2 Department of Materials Science and Engineering, Northwestern University, Evanston, IL 60208, USA*

* Corresponding author: Chuang.Deng@umanitoba.ca

## Abstract

Interfaces are ubiquitous in materials and play a central role in microstructural evolution and material properties. Although interface migration has been studied for more than a century and remains an active field, several foundational assumptions of interface kinetics remain largely untested. In particular, interfaces are commonly treated as massless objects governed by overdamped dynamics. In this study, we show that grain boundaries exhibit measurable inertial behavior under high-frequency oscillatory driving. We introduce a quantitative method to extract an effective interface mass from the phase lag between the applied force and the interface velocity, and find that this mass scales with the atoms participating in boundary migration. Using this framework, we identify regimes in which inertial effects significantly modify interfacial kinetics, especially at frequencies relevant to thermal fluctuations. These results challenge the conventional overdamped description and establish effective interface mass as a key ingredient in a physically complete theory of interface migration.

## Introduction

Interface kinetics plays a central role in materials engineering(1, 2), governing processes such as microstructure evolution(3, 4), solidification(5), phase transition(6, 7), and the development of interface networks(8–10) and textures(11, 12), which influence various material properties(13–17). A growing perspective views the interface as a collective dynamic entity(18–25), whose motion can be described by Langevin equation:

$$m\ddot{x} + \mu\dot{x} = f(t) \tag{1}$$

where, the first and second derivatives of the interface displacement, $\dot{x}$ and $\ddot{x}$ represent the interface velocity and acceleration, *t* is time, $\mu$ is damping coefficient, and $f(t)$ is the general driving force including both stochastic forces such as thermal fluctuations, and deterministic forces such as curvature and energy jump across the interface. The term *m* represents the effective mass of the interface. Traditionally, an interface is viewed as a massless geometric boundary, its motion is therefore modeled using an overdamped Langevin description, where velocity is linearly proportional to the driving force: $\mu\dot{x} = f(t)$. This assumption implies that interfaces respond instantaneously to external forces, neglecting any inertial effects, and has been widely adopted in prevailing interface kinetic models such as the interface random walk theory(18, 26, 27) and various mobility-based frameworks(28–30).

However, since interface migration involves the collective movement of surrounding atoms, their response to external forces cannot always be instantaneous, particularly under high-frequency thermal fluctuations. Ignoring the inertial term $m\ddot{x}$ simplifies mathematics but fails to capture short-time dynamics. Recent theoretical works suggest that this simplification breaks down in the presence of impunities in the grain boundary (GB) (31, 32) and fundamentally alters the symmetry of the interface mobility tensor(24). Despite the frequent use of the overdamped assumption in prior frameworks(18, 24, 27–30, 33), it has never been directly verified due to the challenge of quantifying the effective mass of an interface.

In this study, we propose a method to determine the effective mass of a migrating interface. This approach enables the quantitative inclusion of the inertial term in the Langevin equation, leading to more reliable models of interface kinetics. Moreover, because the effective mass reflects the collective motion of atoms involved in the interface kinetics, our method offers a new perspective for exploring interface migration behavior.

**Simulation and Calculation Methods**

Bicrystal models containing Σ3 (110) GBs were constructed for face-centered cubic (FCC) metals including Al, Ni, Cu, Ag, and Au. The most stable configurations, corresponding to the minimum GB energy, were generated using the method introduced by Olmsted et al.(34). A schematic of the model is shown in Fig. 1a. Periodic boundary conditions were applied along directions parallel to the GB plane. Interatomic interactions were described using embedded atom method (EAM) potentials(35–40). Prior to the production runs, the models were scaled to account for thermal expansion, then equilibrated at target temperature under an NPT ensemble for 200 ps, followed by a brief 10 ps run under NVT ensemble. A Nosé-Hoover thermostat and barostat(41–46) were utilized to control the system temperature and pressure. Subsequent GB migration simulations were conducted under the NVT ensemble. The effect of thermostat is evaluated in Supplemental Materials. The GB position was tracked using the order parameter introduced by Ulomek et al.(47, 48), and its migration velocity was calculated as the average displacement per picosecond. All simulations were performed using LAMMPS(49, 50), and atomic configurations were visualized with OVITO(51). To drive GB migration, a cosine oscillating synthetic energy jump(47, 48) was applied across the boundary, described by

$$f(t) = \frac{1}{2} f_0 \cos(\omega t) + f_0 \tag{2}$$

where, $f_0$ and $\omega$ are the amplitude and angular frequency of the driving force, respectively. As derived in the Appendix, the interface responds with a velocity of the form

$$\dot{x}(t) = -A\omega \sin(\omega t - \delta) + f_0/\mu \quad (3)$$

Here, $A$ is a constant, $\delta$ is the phase parameter of the velocity fit, and the velocity phase lag relative to the cosine drive (Eq. 2) is $\delta' = \delta - \pi/2$. To minimize the effect of thermal noise and accurately fit Eq. 3, the recorded velocity signals were processed using a fast Fourier transform (FFT). Components outside the target frequency band $\omega/2\pi(1 \pm 0.1)$ were filtered out, while the zero-frequency component was retained to obtain the mean interface velocity $\dot{x}_0$. As shown in Fig. 1b, at high loading frequencies (6.283 × $10^{11}$ rad/s), there is a clear phase lag $\delta' = \delta - \pi/2$ between the GB velocity and the applied loading. According to the derivation in Appendix, the fitted phase $\delta$ in Eq. 3 satisfies the following relationship with loading frequency $\omega$:

$$\tan\delta = \frac{-\mu}{m\omega} \quad (4)$$

From Eq. 3, the damping coefficient $\mu$ can be determined from the mean velocity as

$$\mu = \frac{f_0}{\dot{x}_0} \quad (5)$$

In this way, the effective mass of the GB *m* can be extracted by fitting the slope of $\tan\delta$ *vs*. $1/\omega$ curve. An example from Al GB model under various loading frequencies is shown in Fig. 1c, exhibiting excellent agreement with the relationship described by Eq. 4.

To avoid variations in $\mu$ caused by large driving forces(52, 53), a small driving force in the range of $f_0$ =0.001~0.004eV/Ω was applied to measure the effective mass. Here, Ω is average atomic volume obtained through Voronoi analysis(54). Within this range, $\mu$ can be safely regarded as a constant(53). For example, the mobility ($1/\mu$) of the Ni Σ3 (110) GB measured under these conditions ranges from 1.02958×$10^{-5}$ to 1.07211×$10^{-5}$ $m^4$/(J s), which closely matches the value of 1.02×$10^{-5}$ $m^4$/(J s) obtained in the previous study using the interface random walk method at the zero-force limit(24). The same driving conditions were used to determine the effective GB mass for Al, Ni, Cu, Ag, and Au at room temperature (300 K).

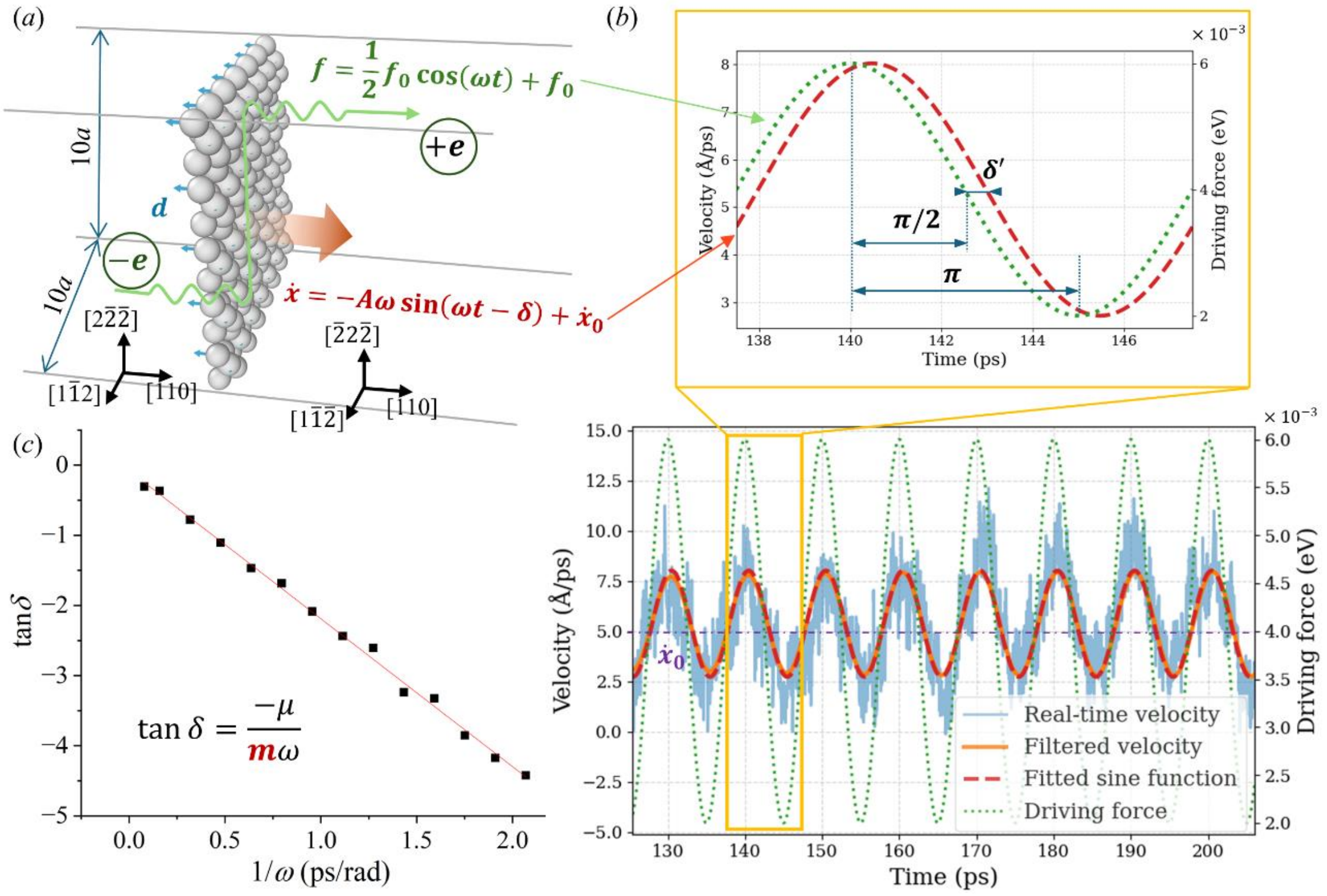

**Figure 1.** (a) Schematic of the simulation model. *a* denotes the lattice parameter, *f* the oscillatory driving force, and ẋ the GB velocity. Blue arrows and *d* indicate perpendicular atomic displacements across the GB. (b) Example of a real-time velocity signal and its FFT-filtered counterpart. The red dotted curve is the fit of filtered velocity to Eq. 3. Data are from an Al Σ3 (110) GB with ω = $6.283\times10^{11}$ rad/s and $f_0$= 0.004 eV/Ω. The data show a clear phase lag $\delta'$ between the interface velocity and the applied loading, where $\delta' = \delta - \pi/2$, and $\delta$ is the fitted phase in Eq. 3. (c) tan $\delta$ versus 1/$\omega$ for the same model and $f_0$ across varying $\omega$.

## Results

Figure 2 shows that the measured GB masses exhibit a strong dependence on the atomic mass of the material, suggesting that the effective GB mass is closely linked to the atoms involved in GB migration. The atomic motion pattern of the Σ3 (110) GB aligns with the conventional atom hopping model, in which atoms on one side of the boundary hop perpendicularly to the other side, driving GB migration, as illustrated in the inset of Fig. 2 based on nudged elastic band(55, 56) method. We calculated the total mass of a single layer of atoms displaced perpendicularly during migration and plotted these values in Fig. 2. The results closely match the measured effective interfacial masses. Although discrepancies between the two curves in Fig. 2 warrant further investigation to identify the specific atoms contributing to the effective GB mass, the consistent trend strongly supports the connection between the effective GB mass and the atomic masses involved in the migration process.

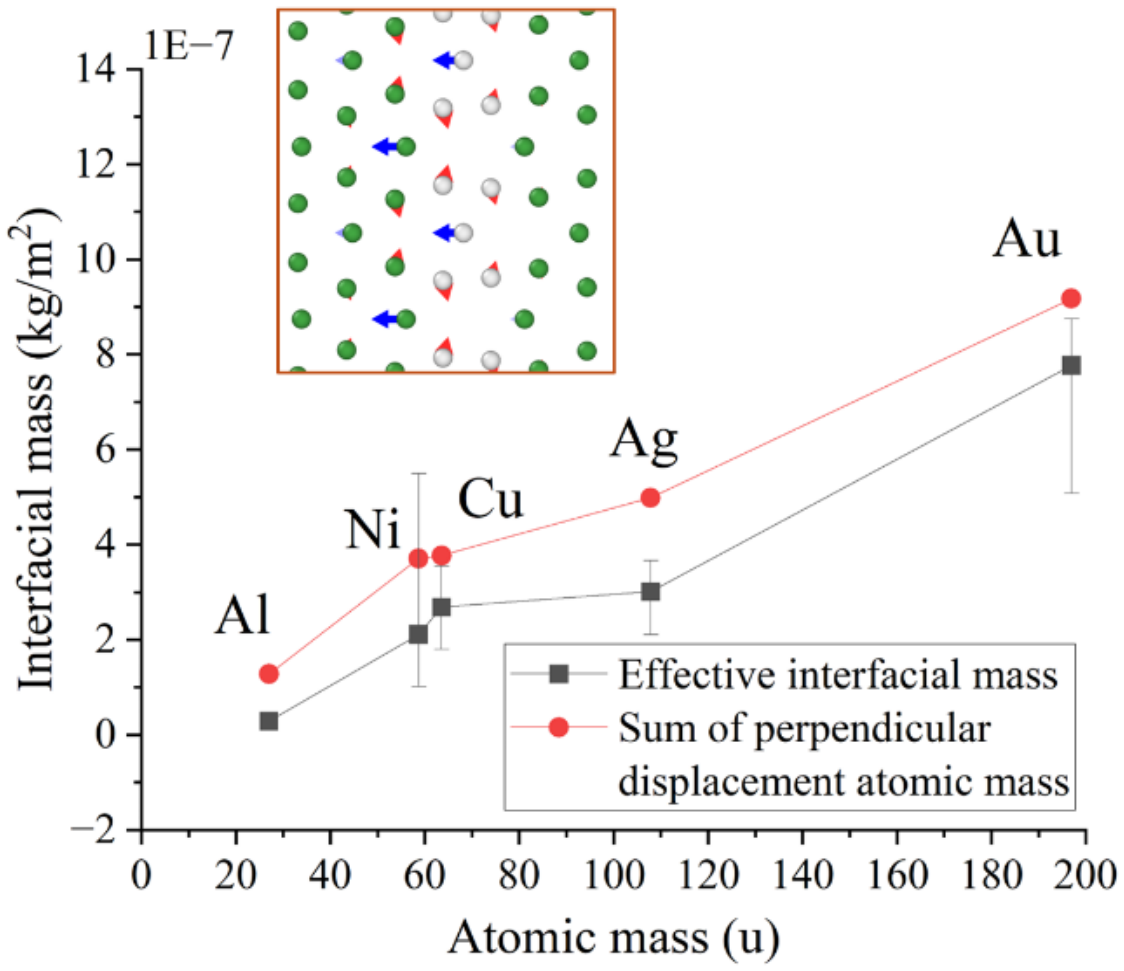

**Figure 2.** Effective GB masses extracted using Eq. 4 for Σ3 (110) GBs in FCC metals. Error bars indicate the effective mass extracted under different $f_0$. Inset shows the GB migration mechanism, where atoms hop perpendicularly across the boundary. The red curve gives the summed mass of one atomic layer undergoing perpendicular displacement for each material.

Figure 3a shows how the measured effective mass varies with driving force. Aside from one outlier in Ag, the effective GB mass increases monotonically with driving force. This trend supports the idea that the effective mass reflects the number and mass of atoms engaged in migration: higher driving force involves more atoms and raises the effective mass. Moreover, Σ3 (110) GB is known to exhibit anti-thermal migration behavior(24, 57), with mobility that decreases as temperature rises. Figure 3b shows that the effective GB mass follows the same trend. As illustrated in the inset of Fig. 2, GB migration proceeds through cooperative atomic motion. Prior studies(58, 59) indicate that higher temperatures disrupt this cooperativity and reduce mobility. The resulting loss of cooperation could also lower the number of atoms participating in migration, which in turn reduces the effective GB mass.

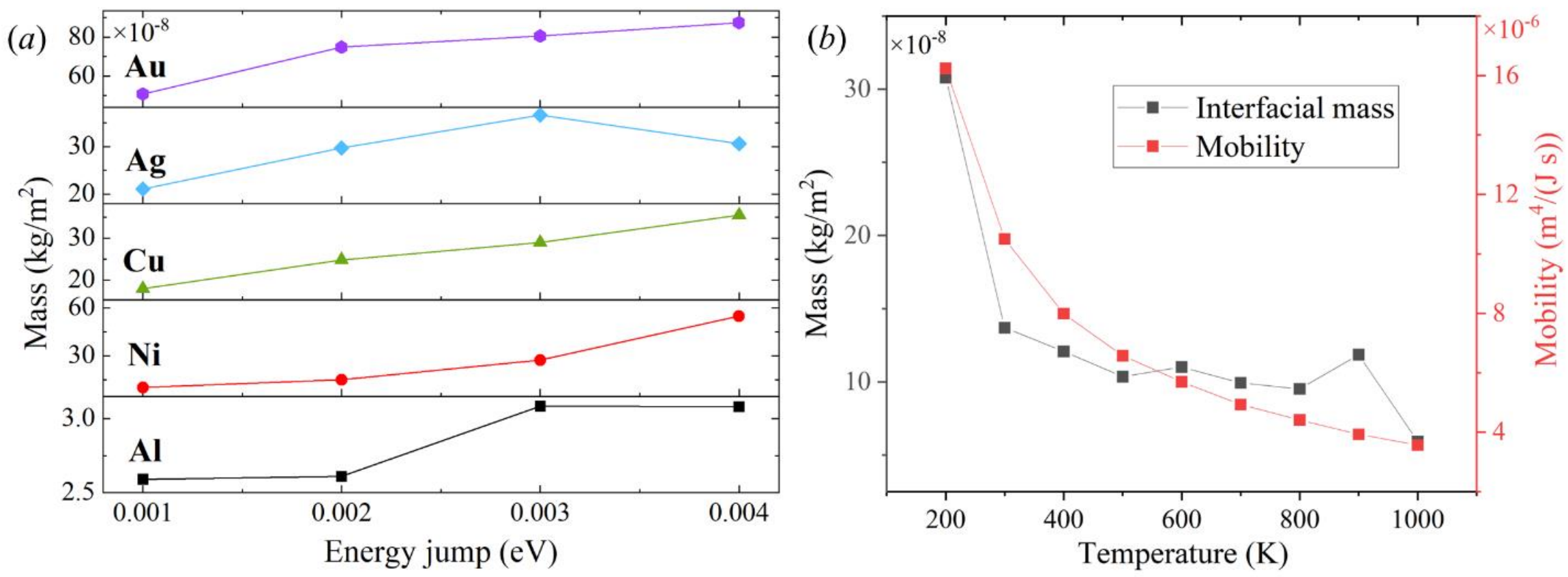

**Figure 3.** (a) Driving force and (b) temperature dependency of the effective GB mass. The mobility data for Ni in panel (b) are from Ref.(24) by interface random walk method.

The effective interface mass plays a key role in evaluating the validity of the "overdamped" kinetics assumption, which has been widely used in previous studies(18, 24, 27–30). To examine this, the effective GB masses obtained were substituted into Eq. 4 to calculate the corresponding phase lag $\delta' = \delta - \pi/2$. As shown in Fig. 4, in the low-frequency regime ($\omega \leq 10^{10}$ rad/s), where the GB velocity is nearly in phase with the applied force ($\delta' \to 0$), as also illustrated in Fig. 1b. In this range, the inertial term in Eq. 1 is negligible. However, as the loading frequency exceeds a certain critical point ($\omega > 10^{10}$ rad/s), $\delta$ rises sharply with the loading frequency, and the velocity becomes progressively out of phase with the driving force ($\delta' \to \pi/2$), indicating that GB kinetics becomes inertia dominated. The yellow band in Fig. 4 marks the Debye frequencies(60) for Al, Ni, Cu, Ag, and Au, which are well above the transition range from in-phase to out-of-phase behavior. This result suggests that under thermal fluctuation or phonon-driven conditions, the inertia term in Eq. 1 cannot be ignored.

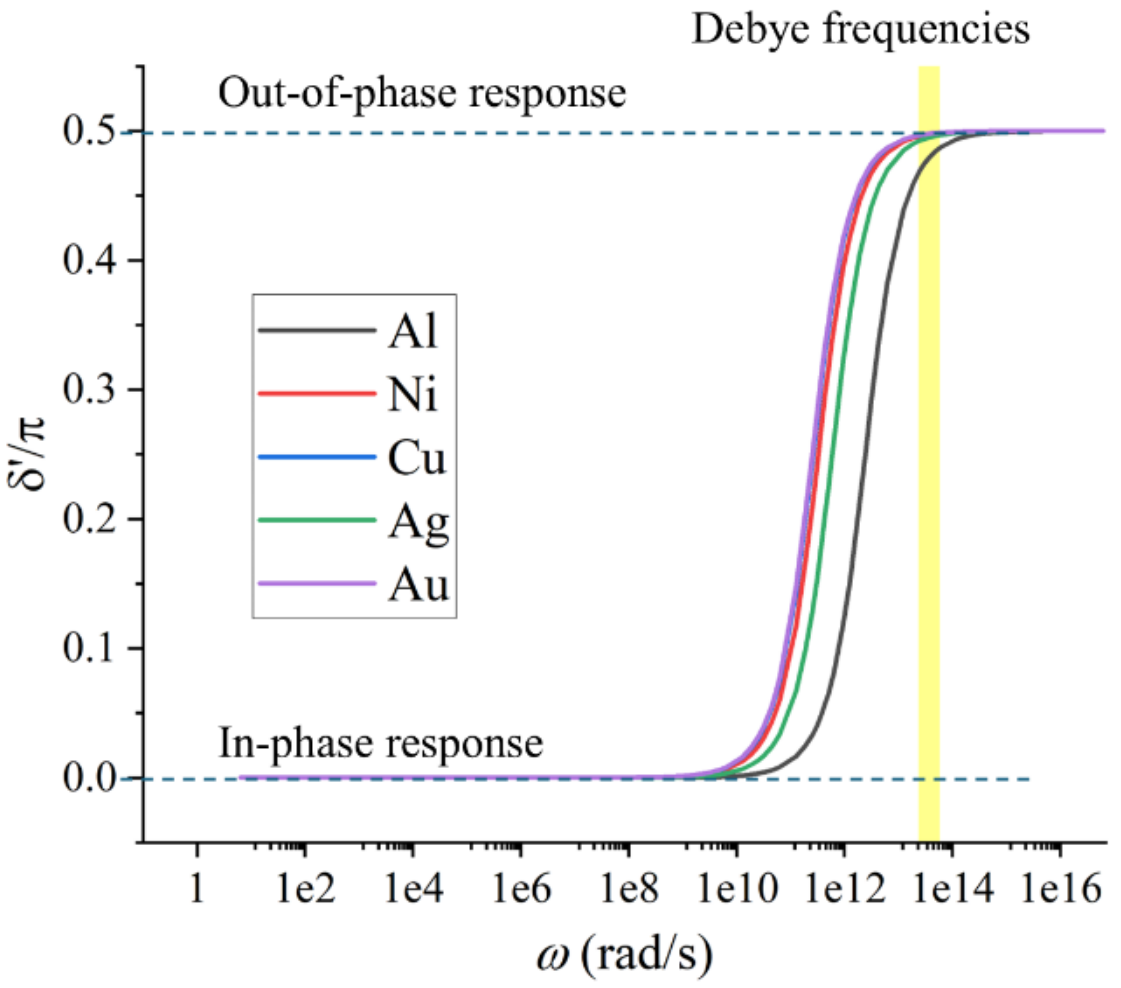

**Figure 4.** Phase lag $\delta' = \delta - \pi/2$ as a function of angular frequency $\omega$. The yellow band marks the Debye frequency range for Al, Ni, Cu, Ag, and Au(60).

**Discussions**

*Extension to more complicated GBs*

Our primary results and analysis focused on the Σ3[110] GB in FCC crystals, as its structural simplicity allows the effective mass to be easily linked to the physical mass of the moving atoms. However, GBs in real polycrystalline materials are generally much more complex, often exhibiting multilayer atomic structures and complicated migration mechanisms. To evaluate the inertia effect in these structures, we expanded our analysis to include a broader selection of GBs, featuring symmetric and asymmetric tilts, mixed characters, and large coincidence site lattice values (Σ values). To ensure reliable measurements of GB velocity and phase, we chose GBs that demonstrated high mobility in our previous survey(24). Detailed structural information for these GBs is provided in Table S1 of the Supplemental Material. The effective mass of these GBs was measured at 300 K under the oscillating driving force described by Eq. 2 ($f_0$ =0.001~0.004eV/Ω).

For all tested GBs, the out-of-phase to in-phase velocity transition (phase lag $\pi/2 \rightarrow 0$) occurs between driving frequencies of 3.14 $\times 10^{11}$ - 6.28 $\times 10^{12}$ rad/s, indicating a similar inertial response. Figure 5 shows that while there is indeed a structural dependency of the effective mass, but the variation is not significant. This suggests that the inertia effect is not severely affected by the presence of complex, multilayer structures alone.

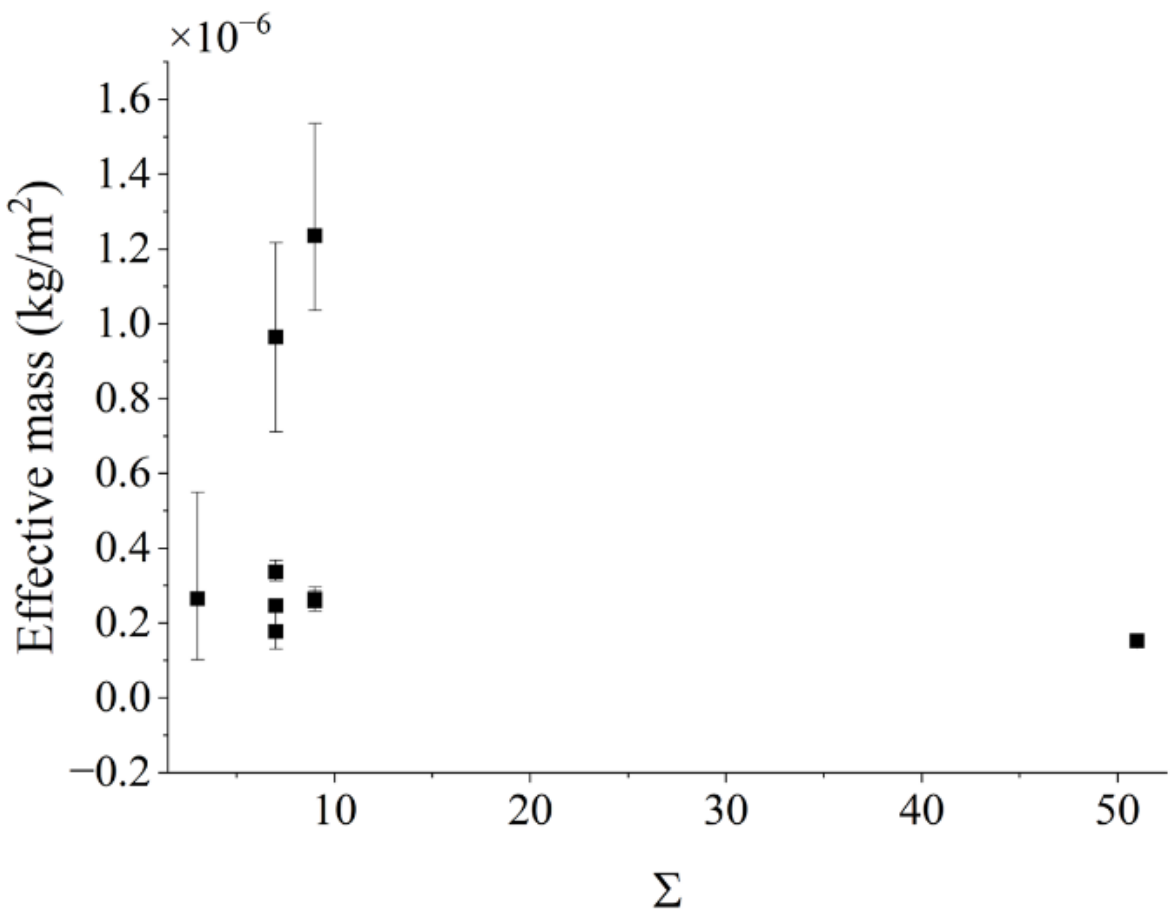

**Figure 5.** Interfacial effective mass as a function of the $\Sigma$ value for Ni GBs at 300K. Structural information for each data point is provided in Table S1 of the Supplemental Material.

Furthermore, GB migration in crystalline materials is usually accompanied by the shear movement of adjacent grains(9, 61). This indicates that the migrating GB must overcome the inertia of not only the interfacial atoms but also the adjacent grain atoms. Consequently, we expect shear-coupled GBs to exhibit a much higher effective mass. We captured the out-of-phase to in-phase velocity transition for the shear-coupled $\Sigma 63\ (10\ 6\ 2)/(10\ 6\ \bar{2})$ GB at a much lower frequency range of 6.28 $\times 10^{9}$ - 6.28 $\times 10^{10}$ rad/s, which demonstrates a much stronger inertial response to the external driving force. The extracted effective mass is $6.54\times 10^{-5}$ kg/m$^2$, which is almost two orders of magnitude higher than that of all non-shear-coupled GBs presented in Fig. 5. This substantial increase confirms that the adjacent grain atoms contribute heavily to the coupled motion, however, this contribution is expected to decrease significantly at elevated temperatures as the shear coupling decays. Beyond shear-coupled migration, pure GB sliding is another critical mechanism governing material deformation(62). Because the mass of the adjacent grains remains the primary contributor during sliding, we anticipate an effective mass of a similar magnitude. However, verifying this requires a more sophisticated method that can exert an immediate shear force directly on the GB and accurately measure the phase lag of the velocity response. Once established, following the mobility tensor framework proposed by Chen et al.(28), these multidimensional dynamics and coupling phenomena can be mathematically captured by treating the effective mass as a tensor, thereby extending the 1D kinetic equation (Eq. 1) to three dimensions.

*Applications*

While the effective line mass of dislocations has been extensively studied(63–65), the concept of effective mass for interfaces, e.g. GBs, has never been formally introduced. Nevertheless, several previous studies have suggested its existence(9, 33, 66–68). For example, disconnections play a key role in governing GB mobility(33) and the evolution of GB networks(9, 68). Since disconnections are also a type of line defect, they should likewise possess an effective mass. Moreover, Deng and Schuh(66) reported a diffusive-to-ballistic transition in GB dynamics with increasing interfacial driving force, where the effective mass may serve as the key parameter linking these two regimes. Furthermore, studies have shown that the mean square displacement (MSD) of GB fluctuations exhibits a nonlinear time dependence at short times, even in pure systems(24, 69, 70). This behavior is characteristic of glass-like relaxation and has been closely associated with the influence of effective mass(71, 72). The work of Sinclair and Rottler(73) further reveals that solute segregation intensifies this nonlinear relaxation behavior. As the effective interfacial mass reflects the collective inertia of atoms participating in interface migration, its quantification could provide a complementary kinetic parameter that can capture the dynamics of mass transport across the interface.

Furthermore, since the current study demonstrates the inertial effects of GBs and the presence of an interfacial effective mass, it raises the intriguing possibility that resonance phenomena can occur within GBs. Earlier theoretical work by Garber and Granato(74) established the fundamental physics of resonance in crystallographic defects. Recent molecular dynamics simulations have shown that localized resonance can induce the structural decomposition of dislocations(75). Li et al.'s work(76) revealed resonance in 2D misfit dislocation networks, which are structurally analogous to GBs. If GBs can similarly achieve resonance, the resulting amplified atomic vibrations could bypass the traditional energetic limits, potentially leading to GB decomposition, structural phase transitions, or the nucleation of disconnections. This suggests a novel pathway for GB engineering: manually triggering these specific processes by applying targeted vibrational frequencies to GBs. The quantification framework introduced in our study provides a clear reference and theoretical foundation to guide this future technique.

## Conclusions

Previous interface kinetics studies have predominantly relied on overdamped assumptions, omitting the mass-related inertia term. In this study, we introduce a method to quantify the effective mass of a migrating interface, with our molecular dynamics simulations demonstrating strong agreement with theoretical predictions. Our analysis reveals that the effective mass of the GB is approximately equivalent to a single atomic layer. While inertial effects can generally be ignored under low-frequency loading, they become significant at high frequencies, such as under phonon driving forces.

Although this work focuses on GBs, the approach is expected to be general and apply to other migrating interfaces such as phase boundaries. Incorporating effective mass into kinetic descriptions should yield more accurate models and refine how we understand and simulate interface migration, and open broader avenues for future research.

## Acknowledgments

This research was supported by NSERC Discovery Grant (RGPIN-2019-05834), Canada, and the use of computing resources provided by Research Alliance of Canada. X. Song acknowledges financial support from US Department of Energy award No. DE-SC0025287. During the preparation of this manuscript the authors used ChatGPT to improve its readability. After using this tool, the authors reviewed and edited the manuscript as needed and take full responsibility for the content of the publication.

## Data Availability

The data and codes that support the findings of this study are available from the corresponding author upon reasonable request.

## Appendix: Theory for Determining Interface Effective Mass from Velocity Response to Oscillatory Forcing

This derivation is adapted from the procedure outlined by Taylor for the motion of a damped particle(77). The motion of an interface subjected to a cosine oscillating force is described by (after filtering out the effect of thermal noise)

$$m\ddot{x} + \mu\dot{x} = f\cos\omega t + f_0 \tag{A1}$$

Here, $\dot{x}$ and $\ddot{x}$, are the interface velocity and acceleration, respectively, *t* is time, $\mu$ is the damping coefficient, *m* is the effective mass of the interface, and $\omega$ is the angular frequency of the applied oscillation. The terms $f$ and $f_0$ denote the amplitudes of the oscillatory and constant driving forces, respectively. Introducing normalized parameters $\beta = \mu/m$, and $f^* = f/m$, $f_0^* = f_0/m$, Eq. A1 reduces to

$$\ddot{x} + \beta\dot{x} = f^*\cos\omega t + f_0^* \tag{A2}$$

Similarly, the motion under a sine oscillating force is expressed as:

$$\ddot{y} + \beta\dot{y} = f^*\sin\omega t + f_0^* \tag{A3}$$

Define the complex function

$$z(t) = x(t) + iy(t) \tag{A4}$$

From Eqs. A2 and A3, we get

$$\ddot{z} + \beta \dot{z} = f^*(\cos \omega t + i \sin \omega t) + f_0^* + i f_0^* \\ = f^* e^{iwt} + f_0^*(1+i) \quad \text{(A5)}$$

Assume a particular solution for Eq. A5

$$z(t) = C_1 e^{iwt} + C_2 t \quad \text{(A6)}$$

where both $C_1$ and $C_2$ are complex constants. Substituting Eq. A6 into Eq. A5 yields

$$C_1(i\omega\beta - \omega^2)e^{iwt} + \beta C_2 = f^* e^{iwt} + f_0^*(1+i) \quad \text{(A7)}$$

Since $e^{iwt} \neq 0$, we must have

$$C_1 = \frac{f^*}{i\omega\beta - \omega^2} \quad \text{(A8)}$$

$$C_2 = f_0^*(1+i)/\beta \quad \text{(A9)}$$

Rewrite the complex coefficient $C_1$ in polar form

$$C_1 = Ae^{-i\delta} \quad \text{(A10)}$$

then

$$A^2 = C_1 C_1^* = \frac{f^*}{i\omega\beta - \omega^2} \cdot \frac{f^*}{-i\omega\beta - \omega^2} = \frac{{f^*}^2}{\omega^2\beta^2 + \omega^4} \quad \text{(A11)}$$

Combining Eqs. A8 and A10, we get

$$A(i\omega\beta - \omega^2) = f^* e^{i\delta} \quad \text{(A12)}$$

Since both $A$ and $f^*$ are real, the phase angle $\delta$ must match the argument of $i\omega\beta - \omega^2$. Therefore,

$$\tan\delta = \frac{\beta}{-\omega} \quad \text{(A13)}$$

The full solution of Eq. A5 is

$$\begin{aligned} z(t) &= C_1 e^{i\omega t} + C_2 t \\ &= A e^{i(\omega t - \delta)} + f_0^*(1+i)t/\beta \quad \text{(A14)} \\ &= A\cos(\omega t - \delta) + f_0^* t/\beta + i[A\sin(\omega t - \delta) + f_0^* t/\beta] \end{aligned}$$

Separating into real and imaginary parts using Eq. A4, we obtain

$$x(t) = A\cos(\omega t - \delta) + f_0^* t/\beta \quad \text{(A15)}$$

$$y(t) = A\sin(\omega t - \delta) + f_0^* t/\beta \quad \text{(A16)}$$

Taking time derivatives and restoring the original parameters $\beta = \mu/m$ and $f_0^* = f_0/m$, we get

$$\dot{x}(t) = -A\omega\sin(\omega t - \delta) + f_0/\mu \quad \text{(A17)}$$

$$\dot{y}(t) = A\omega\cos(\omega t - \delta) + f_0/\mu \quad \text{(A18)}$$

Supplemental Materials

for

**Effective Mass of a Migrating Interface**

Xinyuan Song[1, 2], Chuang Deng[1*]

*1 Department of Mechanical Engineering, University of Manitoba, Winnipeg, MB R3T 2N2, Canada*

*2 Department of Materials Science and Engineering, Northwestern University, Evanston, IL 60208, USA*

* Corresponding author: Chuang.Deng@umanitoba.ca

## S1. The effect of Nose-Hoover thermostat

In the current study, we applied the Nose-Hoover thermostat(1, 2) to simulate the Gibbs's canonical ensemble. This approach regulates the system by introducing a dynamic drag force into the standard equations of motion, modifying the momentum equation to:

$$\dot{\boldsymbol{p}}_i = \boldsymbol{F}_i - \zeta \boldsymbol{p}_i \tag{S1}$$

where, $\dot{\boldsymbol{p}}_i$ is the time derivative of the momentum of particle $i$, $\boldsymbol{F}_i$ is the interatomic force acting on particle $i$, and $\zeta$ is the dynamic friction coefficient. The evolution of $\zeta$ is governed by a feedback loop controlled by the relaxation time $\tau$, which describes how aggressively the thermostat responds to deviations from the target kinetic energy.

To evaluate the influence of the thermostat on our results, we tested the relaxation time, $\tau$, across various timescales: 0.05 ps, 0.1 ps, 0.2 ps (the value used in this study), 0.5 ps, 1 ps, 5 ps, and 10 ps. As $\tau$ increases, the thermostat coupling weakens, causing the system dynamics to more closely approximate a microcanonical (NVE) ensemble. We applied a oscillating driving force to the grain boundary (GB) described by:

$$f(t) = \frac{1}{2} f_0 \cos(\omega t) + f_0 \quad \text{(S2)}$$

We used a base force amplitude of $f_0$= 0.004 eV/Ω, where Ω is atomic volume. Because this represents the maximum driving force applied in our study, it elicits the most aggressive response from the thermostat. For comparison, we also performed simulations under a NVE ensemble, where the work done by the driving force caused the system temperature to increase by approximately 12 K by the end of the simulation. These tests were conducted in an aluminum system across a range of driving frequencies, $\omega$. This setup is identical to the one used for the results shown in Fig. 1c of the main article. Finally, we recorded the resulting velocity phase lag, $\delta$, defined by the equation:

$$\dot{x}(t) = -A\omega \sin(\omega t - \delta) + f_0/\mu \quad \text{(S3)}$$

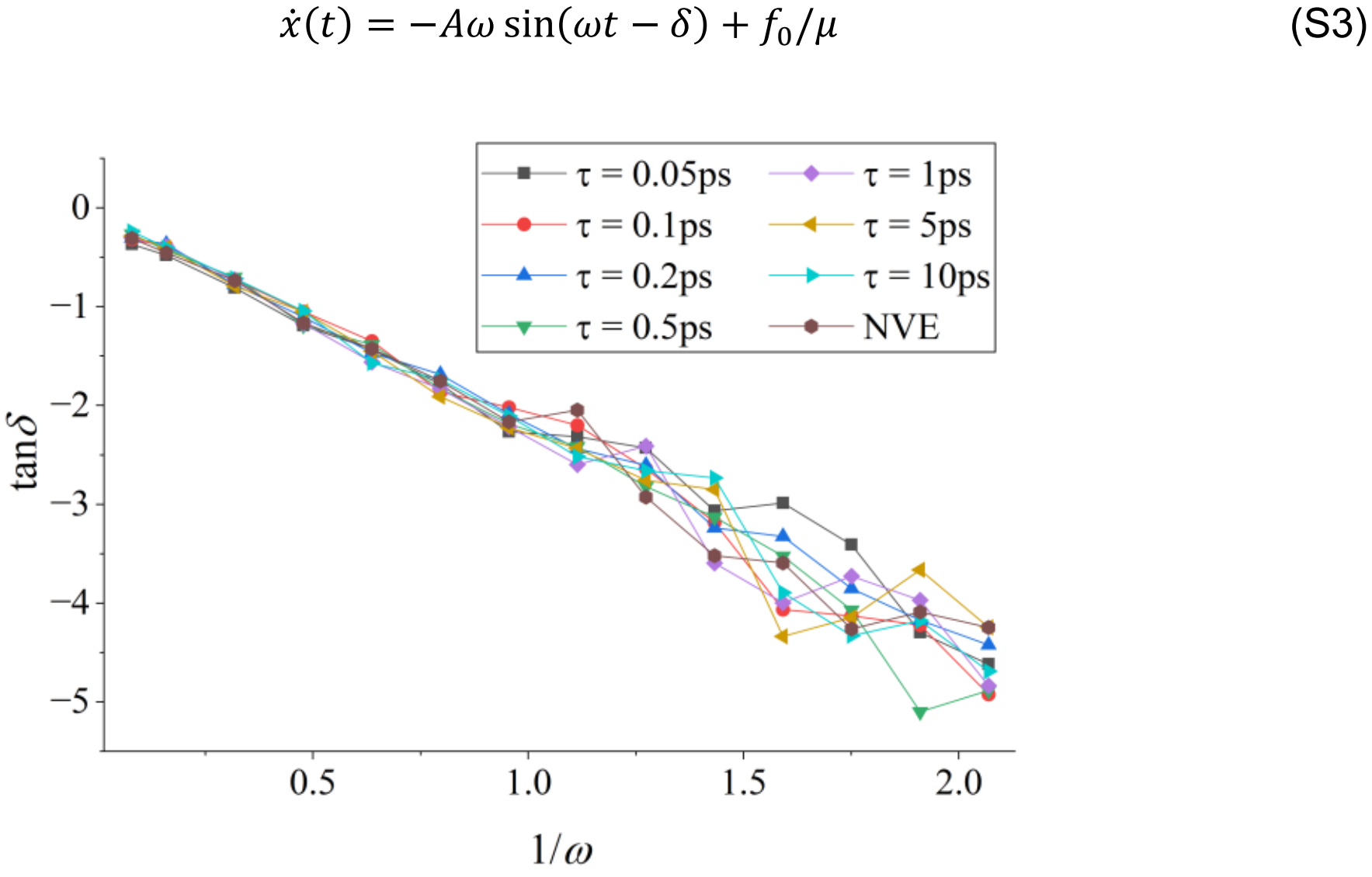

**Figure S1** Comparison of $\tan\delta$ versus 1/ $\omega$ under different thermostat relaxation time $\tau$ and a strict NVE ensemble at 300K.

The simulation results are shown in Fig. S1. At high driving frequencies (low 1/ $\omega$), the extracted $\tan\delta$ values remain consistent across the different $\tau$ (including NVE). However, at lower frequencies (higher 1/ $\omega$), the data fluctuation becomes significant. This variance is likely due to inherent mathematical sensitivity. At lower driving frequencies, the GB has more time to relax, shifting its velocity response from out-of-phase to in-phase with the driving force. This transition causes the phase lag $\delta$ in Eq. R2 to shift from $\pi$ to $\pi/2$. Because $\tan\delta$ approaches an asymptote near $\pi/2$, the calculation becomes extremely sensitive to minor measurement errors in $\delta$, amplifying the statistical noise in this regime.

To verify that this is a mathematical sensitivity rather than a thermostat artifact, we performed multiple simulations using the same thermostat relaxation time ($\tau$ = 0.2ps) but with different initial velocity distributions (Fig. S2). The results demonstrate that even within the exact same ensemble, independent initializations produce similarly large fluctuations in the low-frequency regime. Therefore, we can safely conclude that the artificial drag force introduced by the Nosé-Hoover thermostat (Eq. S1) has a negligible effect on the measurement of the interfacial mass. Furthermore, the fitted interfacial masses extracted from Fig. S1 range only from 2.99 to 3.53 × $10^{-8}$ kg/m$^2$, a minor variation that does not alter any conclusions in the current study.

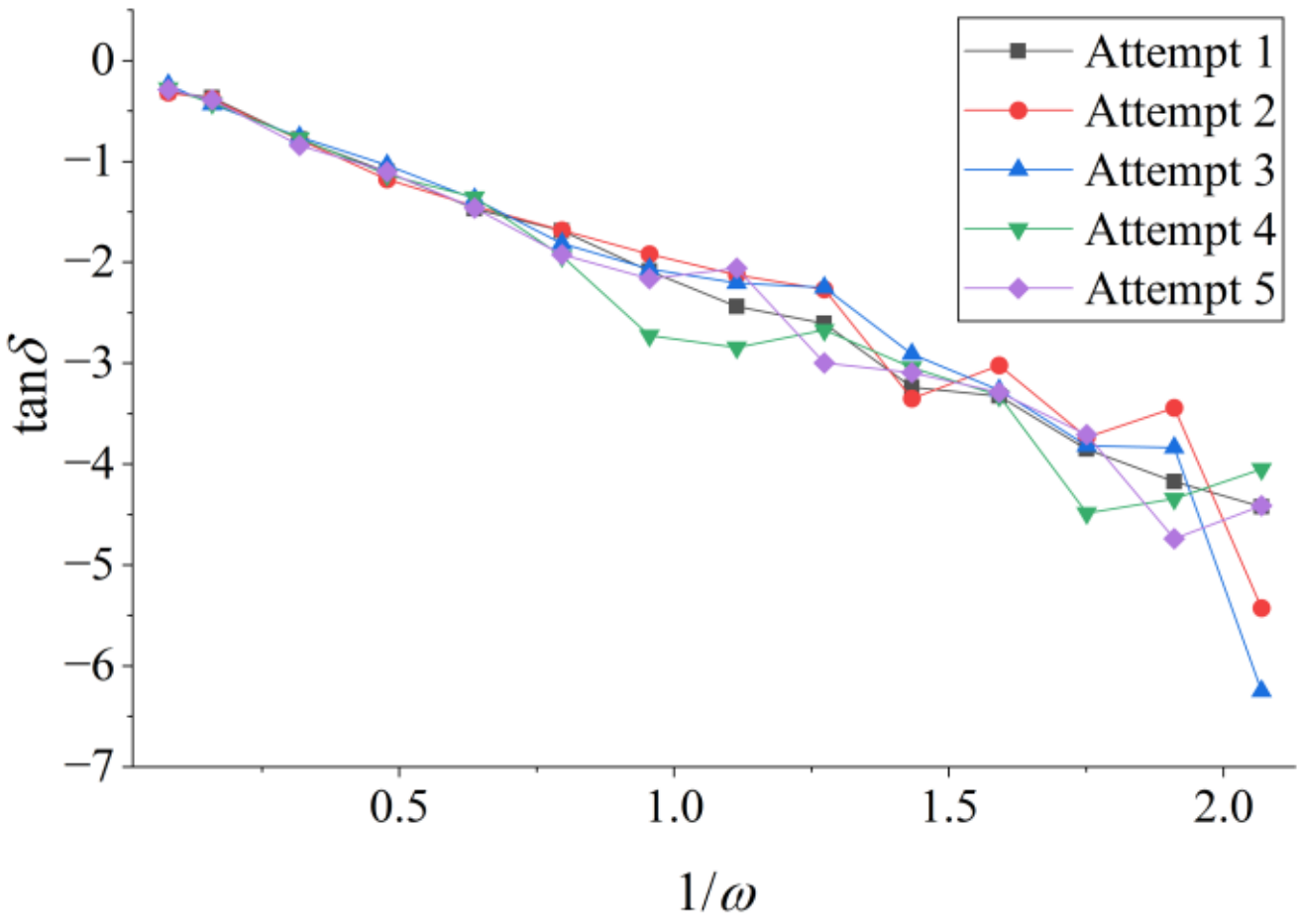

**Figure S2** Comparison of $\tan\delta$ versus 1/ $\omega$ for a fixed thermostat relaxation time $\tau$ = 0.2ps) across five distinct initial velocity seeds at 300K.

## S2. Structural Information of GBs

**Table S1** Detailed structural information and effective interfacial masses of the investigated GBs.

| Σ | GB Character | Orientation Matrix[a] | | Effective interfacial mass (kg/m$^2$) [b] |
|---|---|---|---|---|
| | | Grain 1 | Grain 2 | |
| Non-shear coupled | | | | |
| 3 | tilt | $\begin{bmatrix} 1 & 1 & 0 \\ 1 & \bar{1} & 2 \\ 2 & \bar{2} & \bar{2} \end{bmatrix}$ | $\begin{bmatrix} 1 & 1 & 0 \\ 1 & \bar{1} & \bar{2} \\ \bar{2} & 2 & \bar{2} \end{bmatrix}$ | $2.65\ ^{+2.85}_{-1.63} \times 10^{-7}$ |
| 7 | mixed | $\begin{bmatrix} 4 & 2 & 0 \\ 1 & \bar{2} & 3 \\ 3 & \bar{6} & \bar{5} \end{bmatrix}$ | $\begin{bmatrix} 4 & 2 & 0 \\ \bar{1} & 2 & 3 \\ 3 & \bar{6} & 5 \end{bmatrix}$ | $1.77\ ^{+0.59}_{-0.46} \times 10^{-7}$ |
| 7 | mixed | $\begin{bmatrix} 2 & 1 & 1 \\ 2 & \bar{3} & \bar{1} \\ 2 & 4 & \bar{8} \end{bmatrix}$ | $\begin{bmatrix} 2 & 1 & 1 \\ \bar{2} & 1 & 3 \\ 2 & \bar{8} & 4 \end{bmatrix}$ | $3.36\ ^{+0.32}_{-0.23} \times 10^{-7}$ |
| 7 | mixed | $\begin{bmatrix} 8 & 5 & 1 \\ 2 & \bar{1} & \overline{11} \\ \bar{6} & 10 & \bar{2} \end{bmatrix}$ | $\begin{bmatrix} 7 & 5 & 4 \\ \bar{5} & \bar{1} & 10 \\ 6 & \overline{10} & 2 \end{bmatrix}$ | $9.64\ ^{+2.53}_{-2.53} \times 10^{-7}$ |
| 7 | tilt | $\begin{bmatrix} 8 & 5 & 3 \\ 4 & \bar{1} & \bar{9} \\ \bar{6} & 12 & \bar{4} \end{bmatrix}$ | $\begin{bmatrix} 1 & 1 & 0 \\ 7 & \bar{7} & \bar{9} \\ 0 & 0 & \overline{14} \end{bmatrix}$ | $2.46\ ^{+0.13}_{-0.08} \times 10^{-7}$ |
| 9 | mixed | $\begin{bmatrix} 7 & 5 & 4 \\ 4 & \bar{4} & \bar{2} \\ 1 & 5 & \bar{8} \end{bmatrix}$ | $\begin{bmatrix} 7 & 5 & \bar{4} \\ \bar{4} & 4 & \bar{2} \\ 1 & 5 & 8 \end{bmatrix}$ | $1.24\ ^{+0.30}_{-0.19} \times 10^{-6}$ |
| 9 | mixed | $\begin{bmatrix} 8 & 5 & 1 \\ 2 & \bar{4} & 4 \\ 4 & \bar{5} & \bar{7} \end{bmatrix}$ | $\begin{bmatrix} 8 & 5 & \bar{1} \\ \bar{2} & 4 & 4 \\ 4 & \bar{5} & 7 \end{bmatrix}$ | $2.64\ ^{+0.25}_{-0.13} \times 10^{-7}$ |
| 9 | mixed | $\begin{bmatrix} 7 & 2 & 1 \\ 1 & \bar{4} & 1 \\ 2 & \bar{2} & \overline{10} \end{bmatrix}$ | $\begin{bmatrix} 5 & 5 & 2 \\ \bar{3} & 3 & 0 \\ \bar{2} & \bar{2} & 10 \end{bmatrix}$ | $2.59\ ^{+0.38}_{-0.26} \times 10^{-7}$ |
| 51 | mixed | $\begin{bmatrix} 5 & 2 & 1 \\ 5 & \bar{8} & \bar{9} \\ \bar{2} & 10 & \overline{10} \end{bmatrix}$ | $\begin{bmatrix} 5 & 2 & \bar{1} \\ 5 & \bar{8} & 9 \\ 2 & \overline{10} & \overline{10} \end{bmatrix}$ | $1.52\ ^{+0.06}_{-0.1} \times 10^{-7}$ |
| Shear coupled | | | | |
| 63 | mixed | $\begin{bmatrix} 10 & 6 & 2 \\ 5 & \bar{8} & \bar{1} \\ 1 & 2 & \overline{11} \end{bmatrix}$ | $\begin{bmatrix} 10 & 6 & \bar{2} \\ 5 & \bar{8} & 1 \\ \bar{1} & \bar{2} & \overline{11} \end{bmatrix}$ | $6.84\ ^{+0.30}_{-0.30} \times 10^{-5}$ |

a The first row vector in each matrix represents the crystallographic direction normal to the GB plane.

b The effective interfacial masses were measured at 300K, with $f_0$ = 0.001 ~ 0.004eV/Ω.